\documentstyle[prd,aps,12pt,amssymb]{revtex}
\begin{document}
\draft
\textheight=9.0 in
\textwidth=6.4 in
\tighten
\preprint{RUB-TPII-58/93 \\}
\title{PERTURBATIVE AND NONPERTURBATIVE ASPECTS OF EXCLUSIVE
       PROCESSES \\}
\author{N. G. STEFANIS
\thanks{Invited talk presented at the Workshop on Quantum Field
Theoretical Aspects of High Energy Physics, Kyffh\"auser, Germany,
September 20-24, 1993; to be published in the Proceedings.}
       {\rm and} M. BERGMANN}
\address{
         Institut f\"ur Theoretische Physik II  \\
         Ruhr-Universit\"at Bochum  \\
         D-44780 Bochum, Germany\\
         E-mail: nicos@hadron.tp2.ruhr-uni-bochum.de}
\maketitle
\pacs{}
\section{INTRODUCTION}

The theoretical tools for the description of exclusive processes are the
hard-scattering amplitude which describes the process-dependent
quark-gluon interaction within perturbative QCD and the probability
amplitude for finding the lowest twist quark state in each hadron.
The total amplitude is represented by the convolution of these two
parts~\cite{LB80}, assuming factorization of highly off-shell
or large transverse momentum regions of phase space from regions of low
momenta necessary to form bound states.
Higher-twist components, corresponding to a higher number of partons
(quark-pairs and gluons) are supposed to be suppressed by powers of the
momentum transfer $Q^{2}$.
Recent progress~\cite{LS92,JK93} in Sudakov-suppression techniques
provides support for the conjectured infrared protection of the
perturbative picture.
The focus in this talk will be on recent theoretical developments on
exclusive reactions involving nucleons.

\section{PERTURBATIVE ASPECTS}

The $Q^{2}$-dependence of hadronic wave functions is determined by the
renormalization group equation. To leading order, the dynamical
evolution of the lowest twist hadronic distribution amplitude, which is
the hadronic wave function integrated over transverse momenta up to a
resolution scale $Q^{2}$, is described by the one-gluon exchange kernel
subsumming leading logarithms of ladder graphs. Specifically, the nucleon
distribution amplitude $\Phi_{\text{N}}$ is obtained as the solution to
the integrodifferential equation
\begin{equation}
  \Biggl\{
         Q^{2} \frac{\partial }{\partial Q^{2} } +
                 \frac{3C_{\text{F}}} {2\beta }
  \Biggr\}
  \Phi(x_{\text{i}},Q^{2}) = \frac{C_{\text{B}}} {\beta}\;
  \int_{0}^{1}[dy]\; V(x_{\text{i}},y_{\text{i}})\;
  \Phi(y_{\text{i}},Q^{2})
\end{equation}
[$
  \int_{0}^{1} [dx] \equiv \int_{0}^{1} dx_{1} \int_{0}^{1-x_{1}}
  dx_{2} \int_{0}^{1} dx_{3} \delta (1-x_{1}-x_{2}-x_{3})
$],
where $C_{\text{F}}$ and $C_{\text{B}}$ are the Casimir operators of
the fundamental and adjoint representations of $SU(3)$, respectively,
and $\beta$ is the Gell-Mann and Low function. The leading-order
expression for the integral kernel $V$ has been calculated
in~\cite{LB80}.

To solve the evolution equation (1),
$\Phi_{\text{N}}(x_{\text{i}},Q^{2})$
has to be expanded in terms of the eigenfunctions of the integral
kernel $V$. These eigenfunctions correspond to three-quark operators
which are multiplicatively renormalizable, i.e., to operators with
definite anomalous dimensions~\cite{Pes79}. The latter can be determined
by diagonalizing the evolution equation within some appropriate basis.
To this end, one expresses the solution of (1) in the form
\begin{equation}
  \Phi_{\text{N}}(x_{\text{i}},Q^{2})=
  \Phi_{\text{as}}(x_{\text{i}})\sum_{n=0}^{\infty}B_{\text{n}}
  \tilde \Phi_{\text{n}}(x_{\text{i}})\biggl(
  \frac{\alpha_{\text{s}}(Q^{2})}
  {\alpha_{\text{s}}(\mu^{2})}\biggr)^{\gamma_{\text{n}}},
\end{equation}
where
$\{\Phi_{\text{n}}\}_{=0}^{\infty}$
are orthonormalized eigenfunctions of the interaction kernel within a
truncated basis of Appell polynomials of maximum degree $M$.
$\Phi_{\text{as}}(x_{\text{i}})=120x_{1}x_{2}x_{3}$
is the asymptotic form of the nucleon distribution
amplitude~\cite{LB80}. Because the $\gamma_{\text{n}}$ are positive
fractional numbers increasing with $n$, higher terms in (2) are
gradually suppressed. The Appell polynomials are polynomials of two
independent variables, say $x_{1}$ and $x_{3}$. Thus one can expand
$\{\tilde \Phi_{\text{n}}\}$
in terms of the polynomial basis
$\{x_{1}^{\text{i}}x_{3}^{\text{j}}\}_{\text{i},\text{j}=0}^{\infty}$:
\begin{equation}
  \tilde \Phi_{\text{n}} = \sum _{\text{i},\text{j}=0}^{\infty}
  a_{\text{ij}}^{\text{n}}\;x_{1}^{\text{i}}\;x_{3}^{\text{j}}.
\end{equation}
Then defining moments
\begin{equation}
  \Phi_{\text{N}}^{(\text{i}0\text{j})}(\mu^{2}) =
  \int_{0}^{1}[dx]\;x_{1}^{\text{i}}\;x_{2}^{0}\;x_{3}^{\text{j}}\;
  \Phi_{\text{N}}(x_{i},\mu^{2}),
\end{equation}
the expansion coefficients $B_{\text{n}}$ can be formally determined by
inverting (2) to obtain
\begin{equation}
  \frac{B_ {\text{n}} (Q^{2})}{\sqrt{N_{\text{n}}}} =
  \frac{\sqrt{N_{\text{n}}}}{120}
  \biggl(\frac{\alpha_{\text{s}}(Q^{2})}
  {\alpha_{\text{s}}(\mu^{2})}\biggr)^{\gamma_{\text{n}}}
  \sum_{\text{i,j}=0}^{\infty}a_{\text{ij}}^{\text{n}}\
  \Phi_{\text{N}}^{(\text{i}0\text{j})}(\mu^{2}).
\end{equation}
An explicit calculation of the expansion coefficients $B_{n}(\mu^{2})$
involves the orthonormalization of polynomials with two variables---a
problem with no unique solution because the procedure depends on the
order in which it is performed.
To determine the anomalous dimensions
$
 \gamma_{\text{n}} =  \Bigl(\frac{3}{2}\frac{C_{\text{F}}}{\beta}
 + 2 \eta_{\text{n}}\frac{C_{\text{B}}}{\beta}\Bigr),
$
one has to compute first the zeros $\eta_{\text{n}}$ in the
characteristic polynomial that diagonalizes the evolution kernel.
Such a program has been carried out in~\cite{BS93c,BS93d} and a complete
eigenfunction basis has been constructed up to polynomial order $9$.
All previous calculations had been restricted to maximum order
$3$~\cite{LB80,HEG92}. It is noteworthy that up to order $7$, we have
performed the diagonalization of the evolution kernel analytically.
Below order $3$, our anomalous dimensions coincide with those computed
by Peskin~\cite{Pes79}; those of order $3$ confirm the recently
published (numerical) estimates of~\cite{HEG92}. Our results are listed
in Table~I; for more details we refer to~\cite{BS93c}. The large number
of computed eigenvalues---a total of $54$ corresponding to $29$
symmetric and $25$ antisymmetric eigenfunctions---enables the evaluation
of a well-defined pattern (Fig.~1). The trend line of this pattern
follows the empirical power law (solid line)
$
 \gamma_{\text{n}} = 0.37\;{\cal O}(n)^{0.565}.
$
There is, certainly, no evidence that the ensuing global behavior of
baryon anomalous dimensions coincides with that of mesons (dotted
lines), as prematurely suggested in~\cite{HEG92}.
The proposed method can be used to consistently generalize the ansatz
for $\Phi_{\text{N}}$ to higher orders by retaining the correct
evolution behavior~\cite{BS93d}.

The calculation of moments of the nucleon distribution amplitude resides
on nonperturbative techniques and will be considered in the next chapter.
\newpage
\vspace{-0.8 true cm}
\begin{table}
\squeezetable
\caption{ Orthogonal eigenfunctions
$\tilde{\Phi}_n(x_1,x_2,x_3)=\sum_{lk}\, a_{kl}^{n}\, x_1^kx_3^l$
of the nucleon evolution equation (represented by the
coefficient matrix $a_{kl}^{n}$ with $a_{kl}^{n} = S_n \, a_{lk}^{n}$;
$a_{22}^{n}=0$ for all n). The normalization is given by
$\int_0^1\, [dx]\, x_1x_2x_3\, \tilde{\Phi}_k(x_i) \tilde{\Phi}_n(x_i) =
(N_n)^{-1}\,  \delta_{kn}$. }
\label{Tpoly}
\begin{tabular}{rc|rccc}
$ n  $&$ M $&$ S_n $&$\gamma_n$&$\eta_n$&$ N_n$ \\
\hline
 $ 0$&$ 0$&$1$&${2\over {27}}$&$-1$&$120$\\
 $ 1$&$ 1$&$-1$&${{26}\over {81}}$&${2\over 3}$&$1260$\\
 $ 2$&$ 1$&$1$&${{10}\over {27}}$&$1$&$420$\\
 $ 3$&$ 2$&$1$&${{38}\over {81}}$&${5\over 3}$&$756$\\
 $ 4$&$ 2$&$-1$&${{46}\over {81}}$&${7\over 3}$&$34020$\\
 $ 5$&$ 2$&$1$&${{16}\over {27}}$&${5\over 2}$&$1944$\\
 $ 6$&$ 3$&$1$&${{115 - {\sqrt{97}}}\over {162}}$&${{-\left( -79 + {\sqrt{97}}
\right) }\over {24}}$&${{4620\,\left( 485 + 11\,{\sqrt{97}} \right) }\over
{97}}$\\
 $ 7$&$ 3$&$1$&${{115 + {\sqrt{97}}}\over {162}}$&${{79 + {\sqrt{97}}}\over
{24}}$&${{4620\,\left( 485 - 11\,{\sqrt{97}} \right) }\over {97}}$\\
 $ 8$&$ 3$&$-1$&${{559 - {\sqrt{4801}}}\over {810}}$&${{-\left( -379 +
{\sqrt{4801}} \right) }\over {120}}$&${{27720\,\left( 33607 -
247\,{\sqrt{4801}} \right) }\over {4801}}$\\
 $ 9$&$ 3$&$-1$&${{559 + {\sqrt{4801}}}\over {810}}$&${{379 +
{\sqrt{4801}}}\over {120}}$&${{27720\,\left( 33607 + 247\,{\sqrt{4801}} \right)
}\over {4801}}$\\
 $ 10$&$ 4$&$-1$&${{346 - {\sqrt{1081}}}\over {405}}$&${{-\left( -256 +
{\sqrt{1081}} \right) }\over {60}}$&${{196560\,\left( 7567 - 13\,{\sqrt{1081}}
\right) }\over {1081}}$\\
 $ 11$&$ 4$&$-1$&${{346 + {\sqrt{1081}}}\over {405}}$&${{256 +
{\sqrt{1081}}}\over {60}}$&${{196560\,\left( 7567 + 13\,{\sqrt{1081}} \right)
}\over {1081}}$
\end{tabular}
\vspace{-7 pt}
\begin{tabular}{r|cccccccc}
$ n  $&$a_{00}^n$&$a_{10}^n$&$a_{20}^n$&$a_{11}^n$&$a_{30}^n$&
       $a_{21}^n$&$a_{40}^n$&$a_{31}^n$ \\
\hline
 $  0$&$1$&$0$&$0$&$0$&$0$&$0$&$0$&$0$\\
 $  1$&$0$&$1$&$0$&$0$&$0$&$0$&$0$&$0$\\
 $  2$&$-2$&$3$&$0$&$0$&$0$&$0$&$0$&$0$\\
 $  3$&$2$&$-7$&$8$&$4$&$0$&$0$&$0$&$0$\\
 $  4$&$0$&$1$&$-{4\over 3}$&$0$&$0$&$0$&$0$&$0$\\
 $  5$&$2$&$-7$&${{14}\over 3}$&$14$&$0$&$0$&$0$&$0$\\
 $  6$&$1$&$-6$&${{41 + {\sqrt{97}}}\over 4}$&${{3\,\left( 31 - {\sqrt{97}}
\right) }\over 4}$&${{-5\,\left( 17 + {\sqrt{97}} \right) }\over
{16}}$&${{-5\,\left( 31 - {\sqrt{97}} \right) }\over 8}$&$0$&$0$\\
 $  7$&$1$&$-6$&${{41 - {\sqrt{97}}}\over 4}$&${{3\,\left( 31 + {\sqrt{97}}
\right) }\over 4}$&${{-5\,\left( 17 - {\sqrt{97}} \right) }\over
{16}}$&${{-5\,\left( 31 + {\sqrt{97}} \right) }\over 8}$&$0$&$0$\\
 $  8$&$0$&$1$&$-3$&$0$&${{601 + {\sqrt{4801}}}\over {264}}$&${{59 -
{\sqrt{4801}}}\over {44}}$&$0$&$0$\\
 $  9$&$0$&$1$&$-3$&$0$&${{601 - {\sqrt{4801}}}\over {264}}$&${{59 +
{\sqrt{4801}}}\over {44}}$&$0$&$0$\\
 $ 10$&$0$&$1$&$-5$&$0$&${{379 + {\sqrt{1081}}}\over {48}}$&${{61 -
{\sqrt{1081}}}\over 8}$&${{-\left( 159 + {\sqrt{1081}} \right) }\over
{40}}$&${{-\left( 61 - {\sqrt{1081}} \right) }\over 8}$\\
 $ 11$&$0$&$1$&$-5$&$0$&${{379 - {\sqrt{1081}}}\over {48}}$&${{61 +
{\sqrt{1081}}}\over 8}$&${{-\left( 159 - {\sqrt{1081}} \right) }\over
{40}}$&${{-\left( 61 + {\sqrt{1081}} \right) }\over 8}$\\
\end{tabular}
\end{table}

\begin{figure}
\vspace{7.0 true cm}
\caption{Eigenvalues of the evolution equation for symmetric
         (left-hand side) and antisymmetric eigenfunctions
         (right-hand side).}
\end{figure}

\section{NONPERTURBATIVE ASPECTS}

The derivation of the nucleon distribution amplitude from QCD is
intimately connected with confinement and employs nonperturbative
methods, e.g., QCD sum rules~\cite{CZ84a}, lattice gauge
theory~\cite{RSS87} or the direct diagonalization of the light-cone
Hamiltonian within a discretized light-cone setup~\cite{BP91}.
To determine the moments
$
 \Phi_{\text{N}}^{(n_{1}n_{2}n_{3})},
$
a short-distance operator product expansion is performed at some
spacelike momentum $\mu^{2}$ where quark-hadron duality is
valid~\cite{CZ84a}. One considers
($z$ is a lightlike auxiliary vector with $z^{2}=0$)
\begin{eqnarray}
  \Biggl(iz\cdot \frac{\partial}{\partial z_{\text{i}}}
  \Biggr)^{n_{\text{i}}}
  \Phi_{\text{N}}(z_{\text{i}}\cdot p)\Bigg\vert_{z_{\text{i}=0}}
         & = &
   \prod_{\text{i}=1}^{3}
  \Biggl(iz\cdot \frac{\partial}{\partial z_{\text{i}}}
  \Biggr)^{n_{\text{i}}}
  \int_{0}^{1}[dx]\ e^ { -i\sum_{\text{i}=1}^{3}(z_{\text{i}}\cdot
  p)x_{\text{i}} }
  \Phi_{\text{N}}(x_{\text{i}}) \Bigg\vert_{z_{\text{i}=0}}
  \nonumber \\
         & = &
   (z\cdot p)^{n_{1}+n_{2}+n_{3}}
  \Phi_{\text{N}}^{(n_{1}n_{2}n_{3})}
\end{eqnarray}
and evaluates correlators of the form~\cite{COZ89a,KS87}
\begin{eqnarray}
  I^{\,(n_{1}n_{2}n_{3},m)}(q,z) & = & i\int_{}^{}
  d^{4}x \, e^{iq\cdot x}
  <\Omega\vert T\bigl (O_{\gamma}^{\,(n_{1}n_{2}n_{3})}(0)
  \hat O_{\gamma\prime}^{\,(m)}(x)\bigr )\vert\Omega>(z\cdot \gamma)_
  {\gamma \gamma\prime} \nonumber\\
  & & \\
  &  = &
  (z\cdot q)^{\,{n_{1}+n_{2}+n_{3}+m+3}}
  I^{\,(n_{1}n_{2}n_{3},m)}(q^{2})\;,
\nonumber
\end{eqnarray}
where the factor $(z\cdot \gamma )_{\gamma\gamma\prime}$ serves to
project out the leading twist structure in the correlator, and
\begin{equation}
  O^{(n_{1}n_{2}n_{3})} =
  (z\cdot p)^{-(n_{1}+n_{2}+n_{3})}
  \prod_{i=1}^{3}(iz\cdot {\partial \over {\partial
  z_{\text{i}}}})^{n_{\text{i}}}
  O(z_{\text{i}}\cdot p)\big\vert _{z_{\text{i}}=0}
\end{equation}
are appropriate three-quark operators containing derivatives.
Their matrix elements
\begin{equation}
  <\Omega\vert O_{\gamma}^{(n_{1}n_{2}n_{3})}(0)\vert P(p)> =
  f_{N}(z\cdot p)^{\,{n_{1}+n_{2}+n_{3}+1}}N_{\gamma}\,
  O^{\,(n_{1}n_{2}n_{3})}
\end{equation}
are related to moments of the covariant distribution
amplitudes~\cite{HKM75} $V$, $A$, and $T$:
$
 \Phi_{\text{N}}(x_{\text{i}})=V(x_{\text{i}})-A(x_{\text{i}}),
$
$
 \Phi_{\text{N}}(1,3,2)+\Phi_{\text{N}}(2,3,1)=2T(1,2,3)
$
with
V(1,2,3)=V(2,1,3), A(1,2,3)=-A(2,1,3), and T(1,2,3)=T(2,1,3).
Here $f_{\text{N}}$ denotes the ``proton decay constant''.

On the basis of such QCD sum-rule calculations, useful theoretical
constraints on the moments of nucleon distribution amplitudes have been
obtained and various models~\cite{CZ84a,COZ89a,KS87,GS86,SB92a} have
been proposed. Examples of physical observables calculated from these
models are given in~\cite{CZ84a,COZ89a,Ste89} and more recently
in~\cite{SB92a,SB92b,SB93a,SB93b}.
Our project differs from previous ones in that we use a ``hierarchical''
$\chi^{2}$-criterion to parametrize the deviations from the sum-rule
constraints~\cite{BS93a}. This affords to determine optimized versions
of previous model distribution amplitudes as well as to find a new
solution (we labeled heterotic)~\cite{SB92a} which hybridizes
morphological and dynamical features of COZ-type~\cite{COZ89a} and
GS-type~\cite{GS86} amplitudes providing results corroborated by the
available data. The ``hierarchical'' treatment of the sum rules takes
into account the higher stability of the lower-level moments~\cite{Ste89}
and does not overestimate the significance of the still unverified
constraints~\cite{COZ89a} for the third-order moments.
The simple but important assumption is that the model space can be safely
truncated at states with bilinear correlations of fractional momenta
because adding higher-order contributions should only {\it refine} the
initial approximation. The advantage of our method becomes apparent by
taking a more {\it global} approach, i.e., looking for solutions on the
scale of the whole validity range of the sum rules.
This treatmant leads to a characteristic series of local minima shown
in Fig.~2.

\begin{figure}
\vspace{7.7 true cm}
\caption{Large scale pattern of nucleon distribution amplitudes
         complying with existing QCD sum rules.}
\end{figure}

They constitute a pattern characterized by a smooth and finite
{\it orbit} which is completely specified by joint values of $B_{4}$
and
$R\equiv \vert G_{\text{M}}^{\text{n}}\vert / G_{\text{M}}^{\text{p}}$.
This striking scaling behavior seems to pertain to
solutions~\cite{Sch89,HEG92} which incorporate higher-order
eigenfunctions (Appell polynomials)---not used in our fit procedure
(see inset in Fig.~2).
The robustness of the fiducial orbit suggests that many of the features
of nucleon distribution amplitudes that seemed unrelated actually fit
together into a coherent overall structure, hence leveraging our
knowledge of specific characteristics into a more general context.
Isolated samples in the $(B_{4},R)$ plane are relegated to spurious
solutions, either because they exhibit unrealistic large oscillations
in the longitudinal momentum fractions~\cite{Sch89} (stars) or
because they yield a wrong evolution behavior for the nucleon form
factors~\cite{HEG92} (light upside-down triangles).
The presented curves are fits to the local minima of the COZ sum rules
($+$ labels) and to a combined set of KS and COZ sum rules ($\circ$
labels) restricted within the intervals
$0.104\div 0.4881$ and $0.0675\div 0.482$, respectively:
$
 R = 0.437338 - 0.006016 B_{4} - 0.000176 B_{4}^{2}
$
(solid line) and
$
 R = 0.431303 - 0.00752 B_{4} - 0.000241 B_{4}^{2} + 3.851221\times
10^{-6} B_{4}^{3}
$
(dotted line).
Cross-type solutions show up across the fiducial orbit and
they seem to undergo a complete metamorphosis as they pass from COZ-type
solutions ($R=\leq 0.5)$ to the heterotic one (smallest possible value
of $R$ still compatible with the sum-rule constraints). The passage
between these two types of solutions is smooth and the changed shapes
follow an orderly sequence of gradations characterized by comparable
values of $\chi^{2}$. All solutions are organized by the same kind of
dependence between $R$ and $B_{4}$.
The solution denoted $Het^{\prime}$, past the COZ-cluster, has
``mirror''-image characteristics relative to the heterotic
solution~\cite{BS93b,SB93a} but is unstable.

\section{SUMMARY AND CONCLUSIONS}

We have attempted to provide a unified view of nucleon distribution
amplitudes derived on the basis of QCD sum rules.
The pattern that emerges is not one of widely dispersed solutions across
the parameter space, but one of a smooth and finite curve (an orbit) in
the $(B_{4},R)$ plane, as underlined in Figure 2. On the perturbative
side, a systematic theoretical procedure to calculate orthonormalized
eigenfunctions of the leading-order nucleon evolution kernel has been
developed and a complete set of the first $54$ terms has been explicitly
evaluated.

{\it This work was supported in part by the Deutsche
Forschungsgemeinschaft.}

\end{document}